\documentclass[varenna,tikz]{cimento}
\usepackage[latin1]{inputenc}
\usepackage[T1]{fontenc}
\usepackage[english]{babel}
\usepackage{amssymb,amsmath} 
\usepackage{graphicx}
\usepackage{pgfplots} 
\usepackage[mode=buildnew]{standalone}
\usepackage{siunitx}
\usepackage{ulem}
\usepackage{bm}

\title{Two-component spin mixtures}

\author{Giacomo Lamporesi}

\institute{Pitaevskii BEC Center, CNR-INO and Dipartimento di Fisica, Universit\`a di Trento, via Sommarive 14, I-38123 Trento, Italy}

\begin{document}

\maketitle

\begin{abstract}
The high degree of control on ultracold gases allows us to precisely manipulate their internal state.
When the gas is made of atoms in two different internal states, it can be considered as a two-component spin mixture.
Below a critical temperature, the gas becomes a superfluid mixture, never realized before with any other platform, and therefore interesting to study {\it {per se}}, but it also constitutes a promising and versatile platform for applications in spintronic devices or to study phenomena belonging to very different fields, such as magnetism, high-energy physics or gravitation. 
Here, I will revisit ground-state properties and excitations of a binary bosonic superfluid, and then introduce a coherent coupling between the states and treat the global state of the atoms as a spin in the presence of a variable external field.
\end{abstract}


\section{Introduction}
The study of quantum mixtures covers a broad spectrum of physical phenomena. The high degree of tunability of ultracold gases allows us to explore many interesting problems in different configurations, building on the choice of the constituents and the environment where the system is prepared. 

Let us start from the {\bf{constituents}} of the mixture. It is possible to mix together in a single trap atoms of different atomic species (heteronuclear mixtures), with mass ratios ranging from one up to a few tens \cite{Pires2014,Schafer2022}, and very different interaction properties with external fields \cite{Grimm2000}. In such a way, one can mix together atoms with different statistics realizing Bose-Bose, Fermi-Fermi  or even Bose-Fermi mixtures, as reported for instance in Refs. \cite{Mosk2001,Modugno2002}, \cite{Wille2008,Taglieber2008} and \cite{Schreck2001,Hadzibabic2002,Roati2002}, respectively.
A special case of heteronuclear mixture is the isotopic mixture, where  the atomic species is fixed, but different isotopes are involved \cite{Schreck2001,Papp2006,FerrierBarbut2014}. 
One can create mixtures of atoms in different internal spin states \cite{Myatt1997,Stenger1998} of a single atomic species and isotope. 
Differently from heteronuclear mixtures, in spin mixtures all the atoms have equal masses, and state interconversion becomes possible thanks to a proper coupling radiation, that can involve exclusively the internal degrees of freedom \cite{Matthews1999} or realize spin-orbit coupled systems \cite{Lin2011}.


A whole range of interesting physical phenomena, from 
{\bf{few- to many-body}} physics
can be explored.
We can study few-body physics like the collisional scattering properties between distinguishable atoms \cite{Chin2010}, realize artificial molecules \cite{Papp2006,Ni2008,Voigt2009}, Efimov trimers \cite{Ottenstein2008,Pires2014,Tung2014}, or more excited four-body states, as proposed in \cite{Wang2012}.
If the mixture is characterized by a large density imbalance, one can investigate the physics of impurities or the polaron problem \cite{Catani2012,Kohstall2012,Hu2016,Scazza2017}, with isolated particles of one kind immersed in the bulk made of the other. In this case, the bulk can be a fermionic or bosonic gas, and the impurities might be neutral atoms, as well as ions \cite{Zipkes2010,Tomza2019}.
With both components of the mixture being large, one can enter the domain of hydrodynamics \cite{Koschorreck2013,Fava2018,Wilson2021,Cavicchioli2022} and collective phenomena such as phonons  \cite{Kim2020,Cominotti2022}, topological defects \cite{Kang2019,Farolfi2020} and miscibility problems \cite{Miesner1999,Burchianti2020}.

Mixtures can be composed of more than two constituents, both in the case of heteronuclear \cite{Taglieber2008} and homonuclear systems \cite{Ottenstein2008}, realizing exotic compounds or investigating spinor physics \cite{Stenger1998,StamperKurn2013,JimenezGarcia2019,Pagano2014}, for example.
Furthermore, as done for single component quantum gases, mixtures can be studied in different {\bf{geometrical configurations}}, from standard harmonic traps to flat potentials \cite{Beattie2013,BakkaliHassani2021}, in 3D, reduced \cite{Moritz2005,Catani2012,BakkaliHassani2021} or mixed \cite{Lamporesi2010} dimensionality, in the continuum or in lattice geometries \cite{Gunter2006,Ospelkaus2006}.\\

In this Lecture, I will focus on the particular case of a quantum mixture of Sodium atoms in two different spin states. I will start by introducing simple general concepts of Bose-Einstein condensates (BECs) \cite{Stringari2016} and then extend them to the case of two-component spin mixture systems both in the absence and in the presence of coherent coupling between them.

\section{Single component condensate}
\subsection{Ground state BEC}
Let us start by recalling the main basic properties of a weakly interacting bosonic gas that is cooled down to approximately absolute zero temperature. 
A macroscopic fraction of the atoms
occupies the energy ground state and a BEC is formed. In a mean-field approach, it can be generally described using a single wave function $\psi(x,t)=|\psi(x,t)|e^{i\phi(x,t)}$, that obeys the Gross-Pitaevskii equation (GPE) 
\begin{equation}
\label{Eq:GPE}
    i \hbar \frac{\partial }{\partial t} \psi (x,t)= \left(  - \frac{\hbar^2}{2m} \nabla ^2   +V(x,t) +g |\psi(x,t)|^2 \right) \psi(x,t).
\end{equation} 
The stationary ground state solution satisfies 
\begin{equation}
\left(  - \frac{\hbar^2}{2m} \nabla ^2   +V(x) +g |\psi(x)|^2 \right) \psi(x)   = \mu \psi(x).
\end{equation}

Here $\mu$ is the chemical potential, $m$ the atomic mass, $V$ represents the external potential experienced by the gas and $g$ is the contact interaction constant, that can be expressed in terms of the scattering length $a$ as $g=4\pi \hbar^2 a/m$. The interaction constant $g$ can be positive or negative, giving rise to repulsive or attractive condensates ($g=0$ for the ideal gas BEC). It depends on the atomic species and isotope forming the gas, but also on the internal state. The value of $g$ may vary as a function of the external magnetic field, allowing for experimental tunability across Feshbach resonances \cite{Chin2010}.

When $g<0$ the condensate is in general not stable and collapses, unless in case of very small particle number \cite{Donley2001,Pollack2009}. The $g\geq0$ case, instead, provides a stable configuration.  
In the interaction-dominated regime, and in particular for $N \gg 1$ \cite{Dalfovo1999}, the kinetic energy term in Eq.~(\ref{Eq:GPE}) can be neglected -- the so-called Thomas-Fermi (TF) approximation -- leading to a simple relation connecting the actual density distribution in space $n(x)=|\psi(x)|^2$ to the shape of the potential  $V(x)$, given by 
\begin{equation}
    n(x) = \frac{\mu-V(x)}{g}.
\end{equation}
In a harmonic potential $V(x)=\frac{1}{2}m\omega^2x^2$, the typical distribution is therefore well-approximated by an inverted parabola, with a TF radius $R_{TF}=\sqrt{2\mu/m\omega^2}$.

When the BEC is in its ground state, the phase $\phi$ is uniform throughout the sample. If spatial variations of the phase are present, the superfluid locally moves with a velocity 
\begin{equation}
\label{Eq:VelocityField}
    \mathbf{v_s}(x)=\frac{\hbar}{m}\bm{\nabla} \phi(x).
\end{equation}

\subsection{Elementary excitations}
When the ground state system is minimally perturbed, elementary excitations can be introduced. The so-called Bogoljubov spectrum of such quasi particles is given by
\begin{equation}
\label{Eq:Bogoljubov-density}
    E(k) = \hbar \omega(k) = \sqrt{\frac{\hbar^2 k^2}{2m}\left( \frac{\hbar^2 k^2}{2m}+2m\,c^2 \right)} .
\end{equation}
Note that in the limit of small momenta 
$p=\hbar k \ll mc$, the energy is linearly increasing with $k$, with a characteristic slope given by the speed of sound $c$
\begin{equation}
E(k)\simeq \hbar k c  \hspace{3 cm}    c=\sqrt{\frac{ng}{m}},
\end{equation}

while in the opposite limit, $p \gg mc$, we recover the free-particle energy spectrum $E(k)\simeq\hbar^2 k^2/2m$. Any spatial change of the potential locally moving at a velocity $v \ll c$, makes the superfluid adapt to the change and exhibit its non-viscous behavior. In the opposite case, instead, excitations can be emitted like in the Cherenkov regime or 
through the formation of vortical structures.

The interaction energy enhances the possibility of the wave function to show changes within smaller and smaller length scales. A quantity that well expresses such a property is the healing length 
\begin{equation}
    \xi = \frac{1}{\sqrt{8\pi\, n\, a}} = \frac{\hbar}{\sqrt{2m\,n\,g}}.
\end{equation}

Spatial changes of the condensate wavefunction cannot occur on length scales smaller than $\xi$. In fact, topological excitations, such as dark solitons or quantized vortices, that exhibit a phase discontinuity or singularity, have a spatial extension of the order of $\xi$.

\subsection{Topological excitations}

Solitons are localized waves that preserve their features thanks to the competition between dispersion (kinetic energy) and nonlinearity of the GPE (interactions).
Depending on the ratio between the density at the center of the soliton and the background density, solitons have different properties. 
Dark solitons \cite{Burger1999} are a static localized full depletion of a BEC, given by an abrupt phase discontinuity of $\pi$. Solitons can also move inside a BEC (grey solitons). Given their velocity $v_{\rm{sol}}$, all their features are well determined: the phase jump $\Delta \phi$, the spatial length over which the phase jump happens $\Delta x$, and the density depletion $\Delta n$

\begin{equation}
    \Delta \phi=\frac{-2}{\cos{(v_{\rm{sol}}/c)}} \hspace{1cm} \Delta x=\frac{\xi} {\sqrt {1-v_{\rm{sol}}^2/c^2}}  \hspace{1cm} \Delta n=n \left( 1-v_{\rm{sol}}^2/c^2 \right).
\end{equation}

It is easy to verify that faster solitons are wider and show a smaller density depletion than slower ones.
When the density at the center of the soliton is higher than the background one, then one has a bright soliton with a characteristic speed that exceeds the speed of sound. Usually in the Helium community the background density is finite \cite{Ancilotto2018}, but in ultracold gases the term bright soliton is usually referred to localized waves on a zero-density background, typically stabilized by attractive interatomic interactions \cite{Khaykovich2002,Strecker2002}. 

Solitons are stable only in 1D geometries.
In higher dimensions, standard solitons are dynamically unstable and decay into other topological excitations, such as vortex rings, solitonic vortices or vortex dipoles \cite{Mateo2014}.

The relevant quantity that determines the dimensionality of a system, for what concerns the existence and stability of topological defects, is the ratio between the chemical potential and the quantum of the harmonic oscillator energy in the strongly confined directions, $\gamma=\mu/\hbar\omega_\perp$, equivalent to the ratio between the TF radius and twice the healing length, $\gamma=R_\perp/2\xi$. 
In an elongated geometry, if $\gamma \gg 1$, the system is effectively 3D and solitons decay quickly, whereas for  $\gamma\simeq 1$ solitons are stable.\\

Quantized vortices are topologically protected singularities with a $2\pi$ phase winding around the core, and the corresponding atomic flux according to Eq.~(\ref{Eq:VelocityField}). In order to preserve the singlevaluedness of $\psi$, the density drops to zero in the singularities, within a length $\xi$, typical size of the vortex core. Their existence is possible if the system size exceeds $\xi$ in at least two spatial directions.

\section{Two-component spin mixtures}

One particular kind of superfluid mixture that can be experimentally realized consists in having atoms in two different spin states occupying a given volume.
In contrast to heteronuclear and isotopic mixtures, this particular quantum mixture allows for interconversion between the states through the application of a coherent coupling.

In general, we can write the GPEs for each of the two components, that couple thanks to the intercomponent interaction.
In fact, while a single component is characterized by a single interaction constant $g$, in the presence of two components, the relevant interaction constants are three. Let us label the two intracomponent interaction constants $g_a$ and $g_b$ for components $a$ and $b$, and $g_{ab}$ the intercomponent one.
The set of coupled GPEs becomes
\begin{eqnarray}
\label{eq:psia}
     i \hbar \frac{\partial }{\partial t} \psi_{a}(x,t) = \left(  - \frac{\hbar^2}{2m} \nabla ^2   +V(x,t) +g_a |\psi_{a}(x,t)|^2 +g_{ab} |\psi_{b}(x,t)|^2 \right) \psi_{a}(x,t) \\
     \label{eq:psib}
      i \hbar \frac{\partial }{\partial t} \psi_{b}(x,t) = \left(  - \frac{\hbar^2}{2m} \nabla ^2   +V(x,t) +g_b |\psi_{b}(x,t)|^2 +g_{ab} |\psi_a(x,t)|^2\right) \psi_b(x,t).
\end{eqnarray}

with stationary solutions characterized, in general, by two distinct chemical potentials ($\mu_a\neq\mu_b$) for the two components, 
\begin{eqnarray}
 \left(  - \frac{\hbar^2}{2m} \nabla ^2   +V(x) +g_a |\psi_{a}(x)|^2 +g_{ab} |\psi_{b}(x)|^2 \right) \psi_{a}(x) =\mu_a \psi_a(x) \\
 \left(  - \frac{\hbar^2}{2m} \nabla ^2   +V(x) +g_b |\psi_{b}(x)|^2 +g_{ab} |\psi_a(x)|^2\right) \psi_b(x)=\mu_b \psi_b(x) .
\end{eqnarray}

\subsection{Ground state miscibility}
Depending on the values of the three interaction constants, the mixture shows very different behaviors and ground state configurations. 
Let us first focus on the case of a mixture confined in a flat box potential of total fixed volume $V$.
For densities $n_a=|\psi_a|^2$ and  $n_b=|\psi_b|^2$, the energy density of the whole systems can be written as
\begin{equation}
    \mathcal{E}= \frac{1}{2} g_a n_a^2 + \frac{1}{2} g_b n_b^2 + g_{ab} n_a n_b - \mu_a n_a -\mu_b n_b.
\end{equation}
The system is stable and miscible only if the Hessian of $\mathcal{E}$ is positive, and this happens only if all these three conditions are fulfilled
\begin{eqnarray}
\label{eq:ga}  &  &\ \ \ \ \ g_a > 0 \\
\label{eq:gb}   \left [\frac{ \partial ^2 \mathcal{E} }{\partial n_a \partial n_b} \right] > 0 & \ \ \ \ \ \Longleftrightarrow & \ \ \ \ \ g_b > 0 \\
\label{eq:gab}   & & \ \ \ \ \ g_a g_b - g_{ab}^2 > 0
\end{eqnarray}

Different kinds of instabilities can be observed, instead, when either of the above conditions is not satisfied:
\begin{itemize}
    \item If $g_a<0$, the gas of $a$-particles collapses.
    \item If $g_b<0$, the gas of $b$-particles collapses.
    \item If both $g_a$ and $g_b$ are positive, but $g_a g_b < g_{ab}^2$, the two gases become immiscible and undergo phase separation (when $g_{ab}>0$) or collapse together (when $g_{ab}<0$). In this latter case, including also the contribution of beyond mean field effects \cite{Petrov2015}, a further situation can be observed if the intercomponent attraction is small enough and balances the intracomponent repulsion: the system forms self-trapped states known as quantum droplets \cite{Cabrera2017,Semeghini2018,Derrico2019}. 
\end{itemize}

\begin{figure}[t!]
\centering
\includegraphics[width=0.8\columnwidth]{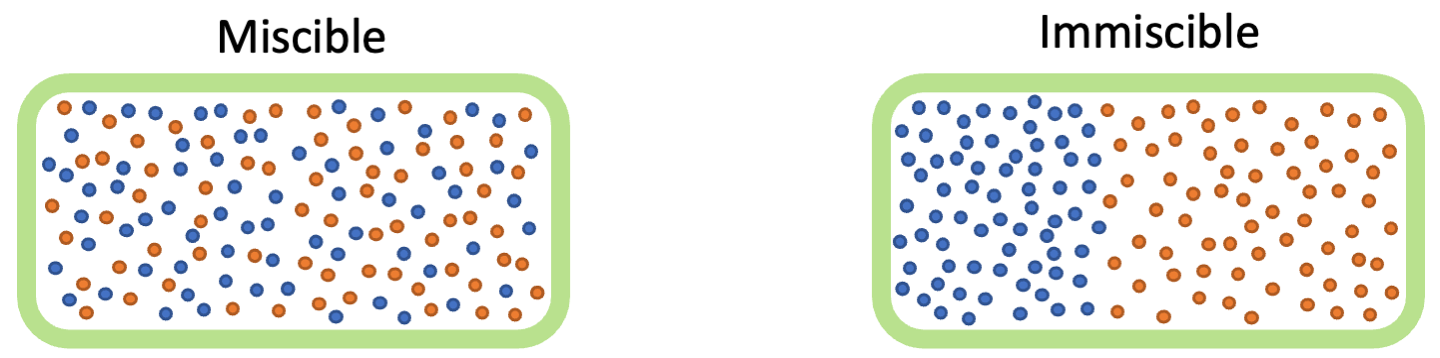}
\caption{Pictorial illustration of the ground-state spatial arrangement of a two-component mixture in a flat-box trap in the miscible (left) or immiscible (right) case. In the first case, both components occupy the whole volume $V$, whereas in the second case $a$-particles (blue) are restricted in a volume $V_a$ and the remaining $V_b=V-V_a$ is occupied by $b$-particles.}
\label{fig:fig-Miscibility}
\end{figure}

In the following, we will restrict to the case where all three interaction constants are repulsive. Then, the two gases will either mix, or undergo a spatial phase separation. 
Intuitively, when the intracomponent interactions are stronger than the intercomponent one, the two gases can spatially overlap (miscible mixture), while phase separation, with different components occupying different spatial regions of volumes $V_a$ and $V_b$ ($V=V_a+V_b$), occurs when the intercomponent interaction dominates (immiscible mixture), as illustrated in Fig.~\ref{fig:fig-Miscibility}. 
If the mixture is in equilibrium, one can estimate \cite{Stringari2016} the energy of the system in the two configurations, miscible ($E_m$) and immiscible ($E_i$), as
\begin{eqnarray}
\label{Eq:Em}
E_m &= &\frac{1}{2} g_a \frac{N_a^2}{ V} +  \frac{1}{2} g_b \frac{N_b^2}{V} + g_{ab} \frac{N_a N_b}{V}  \hspace{2cm} \\
\label{Eq:Ei}
E_i = \frac{1}{2} g_a \frac{N_a^2}{V_a} +  \frac{1}{2} g_b \frac{N_b^2}{V_b} &=& \frac{1}{2} g_a \frac{N_a^2}{V} + \frac{1}{2} g_b \frac{N_b^2}{V} + \sqrt{g_a g_b}\frac{N_a N_b}{V}
\end{eqnarray}

that lead to the following requirement in order to have a stable miscible mixture (miscibility criterium)
\begin{equation}
    \sqrt{g_ag_b}-g_{ab} > 0 .
    \end{equation}
The total density is uniform in the miscible case, but in general, this is not true for the immiscible case, where the regions occupied by the most repulsive component are more dilute. Note, also, that in Eq.~(\ref{Eq:Ei}) we neglected a term associated to the surface tension of the small region that delimits the different domains, where the two gases partially overlap. If one includes such a term as well, the delimiting surface will be the smallest possible, given the trap configuration.

\subsection{Harmonic confinement and buoyancy}
\begin{figure}[t!]
\centering
\includegraphics[width=\columnwidth]{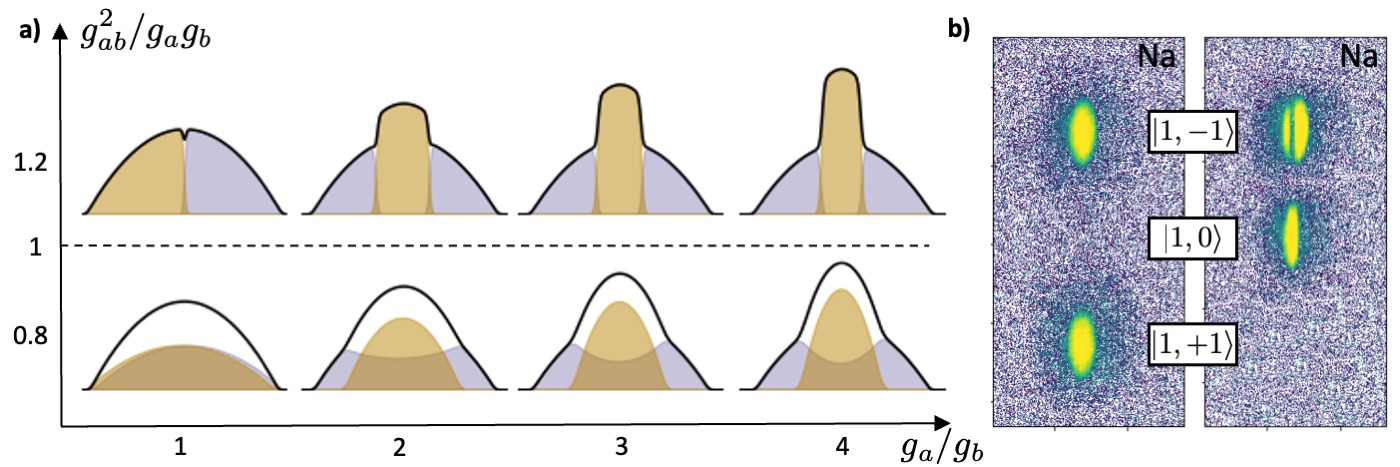}
\caption{ \textbf{a)} GPE simulation of the ground state spatial arrangement of a two-component balanced mixture in a harmonic potential. The shaded yellow and violet regions show the density of each component for different combinations of interaction constants. The black line is the total density. All simulations include a (realistic) tiny magnetic field gradient, added to break the left-right symmetry. \textbf{b)} Experimental observation of a miscible (left) or immiscible (right) Sodium spin mixture, after Stern-Gerlach vertical separation in time of flight.}
\label{fig:fig-Buoyancy}
\end{figure}

When the mixture is confined in a nonuniform trapping potential, the total density is position-dependent and the density distribution of each component is strongly connected to all the interaction constants and to the shape of the confinement. In the commonly used harmonic traps, for instance, even in the miscible case, the two components may occupy the available volume in different ways, one from the other \cite{Ho1996}. If $g_a<g_b$, the minimal energy configuration sees $a$-particles predominantly accumulating in the center of the trap, where the total density is higher, while the more repulsive $b$-particles mainly lying in the outer regions. However, in this case, there is no abrupt interface between them, as in the immiscible case.
Such a buoyancy effect in nonuniform traps can be avoided only if  $g_a=g_b$.
Figure~\ref{fig:fig-Buoyancy}(a) shows the GPE simulation for the spatial distributions of the two components as $g_a/g_b$ is varied, for a miscible case ($g_{ab}^2=0.8\, g_a g_b$) and for an immiscible one ($g_{ab}^2=1.2\, g_a g_b$). 
Note that the total density (black line) no longer follows a TF profile and deviates more and more from it, the larger the difference between $g_a$ and $g_b$.
Two experimental examples, obtained on different combinations of Sodium atoms in the hyperfine $F=1$ manifold, $m_F=\pm1$ (left) and  $m_F=0,-1$ (right), are shown in Fig.~\ref{fig:fig-Buoyancy}(b). Absorption images are taken after a time of flight in the presence of a magnetic field gradient that spatially separates the different components. 
For the $F=1$ spinor of Sodium, the relevant scattering lengths are $a_{+1,+1}=a_{-1,-1}=a_{-1,0}=a_{+1,-0}=54.5~a_0$, $a_{0,0}=52.6~a_0$ and $a_{+1,-1}=50.75~a_0$. Therefore, each two-component mixture of 0 and either $+1$ or $-1$ has $g_a/g_b=1.036$ and $g_{ab}^2/g_a g_b=1.074>1$ (immiscible), whereas the mixture of $+1$ and $-1$ has $g_a/g_b=1$ and $g_{ab}^2/g_a g_b=0.867<1$ (miscible).

\subsection{Density and spin excitations}
In the previous Sections, we saw how the ground states of miscible and immiscible mixtures look like. Especially in the case of miscible mixtures, where an extended spatial overlap is present, it is natural to wonder about elementary excitations. 

The Bogoljubov excitation spectrum for a two-component mixture, is characterized by two branches: the density ($d$) and the spin ($s$) ones. 
They have a similar qualitative behavior,
\begin{equation}
    E_{d,s}(k) = \hbar \omega_{d,s}(k) = \sqrt{\frac{\hbar^2 k^2}{2m}\left( \frac{\hbar^2 k^2}{2m}+2mc_{d,s}^2 \right)},
    \label{Eq:Bogoljubov-ds}
\end{equation}
where density and spin speeds of sound are
\begin{equation}
    c^2_{d,s}= \frac{(g_an_a + g_b n_b) \pm \sqrt{(g_an_a-g_bn_b)^2+4 g_{ab}^2 n_a n_b }}{2m},
    \label{eq:cs-cd}
\end{equation}
where the $+$ is for the density channel and the $-$ for the spin one.
In general, a two-component mixture is characterized by two distinct healing lengths, representing the limiting smallest length scales below which  density or spin excitations cannot be excited
\begin{equation}
    \xi_{d,s}=\frac{\hbar}{\sqrt{2\,}mc_{d,s}}.
       \label{eq:xis-xid}
\end{equation}

Equation~(\ref{eq:cs-cd}), and consequently Eq.~(\ref{eq:xis-xid}), do not provide an intuitive picture of what physically matters in the determination of the characteristic speeds of sound and healing lengths in the system. However, if we consider the special case of a balanced ($n_a=n_b=n/2$) mixture, 
with equal intracomponent interactions ($g=g_a=g_b$),
they simplify to
\begin{equation}
     c^2_{d}= \frac{n}{2m}(g + g_{ab}) \hspace{3cm} c^2_{s}= \frac{n}{2m}(g - g_{ab}),
\end{equation}
and
\begin{equation}
     \xi_{d}= \frac{\hbar}{\sqrt{mn(g + g_{ab})}} \hspace{3cm}  \xi_{s}= \frac{\hbar}{\sqrt{mn(g - g_{ab})}},
\end{equation}
clearly highlighting the different scaling with the combination of interaction constants.
These two sound modes were experimentally observed in \cite{Kim2020}.
We can also introduce the density and spin chemical potentials as 
\begin{equation}
\label{Eq:mu}
     \mu_{d}= n\frac{(g + g_{ab})}{2} \hspace{3cm} \mu_{s}= n \frac{(g - g_{ab})}{2} .
\end{equation}

It is easy to see that, in this particular condition of balanced miscible mixtures with equal intracomponent interactions, the ratio between density and spin quantities shows the exact same scaling, $(c_s/c_d)^2 = \mu_s/\mu_d = (\xi_d/\xi_s)^2 = (g-g_{ab})/(g+g_{ab})$.\\

In the symmetric case, the density and spin channel are also referred to as the total density ($n_a+n_b$) and the magnetization ($n_a-n_b$), respectively.

\subsection{Topological excitations}
In two-component mixtures, one can have topological defects
both in the total density and in the spin channel. In 1D, for instance, when the two components exhibit the same localized phase jump and density depletion, then we have a density soliton, with the properties of solitons in a single component condensate. One can also have an overall flat total density (or at most a small density variation), but a localized solitary wave in the magnetization, with spin-up particles showing a density peak in the density where spin-down ones exhibit a dip. These are the so-called dark-antidark solitons \cite{Kevrekidis2003,Katsimiga2020}, that in the special case of $g_a=g_b \gtrsim g_{ab}$ become practically pure excitations in the spin channel accompanied by a flat total density, and are known as "magnetic solitons" \cite{Qu2016}. A peculiar aspect of such topological structures is that, differently from their counterpart in the density channel, they exhibit a velocity-independent $\pi$ phase jump in the relative phase $\phi$.
Solitons were recently produced also in spinor condensates \cite{Lannig2020,Chai2021}.

Analogously, it is possible to have different kinds of vortices in two-component mixtures. 
Vortices can be in the total density, when both components exhibits a vortex in a well-defined position, with overlapping cores. In this case the characteristic vortex core size is $\xi_d$.
Alternatively, one can have vortices in one component and the other simply filling the density depletion with a localized density peak. In the special case of equal background densities, the latter are known as  half vortices \cite{Gallemi2018}. Half vortices are topological excitations in the magnetization, therefore their spatial extension is $\xi_s$.

\subsection{Sodium two-component mixtures}
In the following, I will show some experimental results obtained using either one of the two spin mixtures of Sodium atoms reported in Table~\ref{tab:Na-interactions}. 
The first one involves two symmetric magnetic states ($m_F=\pm 1$) of the $F=1$ hyperfine manifold. Such a mixture is perfectly miscible and presents no buoyancy. For this reason it is an ideal platform for a clean investigation of spin dynamics and excitations. 
In the second one, atoms in two (extreme) spin states belonging to different hyperfine states $|F,m_F\rangle = |2,-2\rangle$ and $|1,-1\rangle$ are combined in a single trap. These spin components do not mix, but in the presence of a coherent coupling between them, the miscibility can be restored. Such a platform allows for the study of dissipationless ferromagnetic features in atomic superfluids. 

\begin{table}[h]
  \centering
\begin{tabular}{||l||r|r|}
\hline
                &    MISCIBLE               &    IMMISCIBLE \\     
\hline
component $a$   &      $|1,+1\rangle$       &     $|2,-2\rangle$  \\  
component $b$   &      $|1,-1\rangle$       &     $|1,-1\rangle$  \\ 
\hline
$a_a$           & $54.5\,a_0$               &  $64.3\,a_0$ \\
$a_b$           & $54.5\,a_0$               &  $54.5\,a_0$\\
$a_{ab}$        & $50.75\,a_0$              &  $64.3\,a_0$ \\
\hline
$\sqrt{a_a a_b}-a_{ab}$   & $+3.75\,a_0$   &  $-5.1\,a_0$ \\
$\bar{a}=(a_{a}+a_{b})/2$   & $54.5\,a_0$   &  $59.4\,a_0$ \\
$\delta{a_1}=(a_{a}-a_{b})$   & $0\,a_0$      &  $9.8\,a_0$\\
$\delta a_2=(\bar{a}-a_{ab})$   & $+3.75\,a_0$   &  $-4.9\,a_0$ \\
$|\bar{a}-a_{ab}|/(\bar{a}+a_{ab})$   & 3.6 \%    &  4.0 \% \\
\hline
\end{tabular}
    \caption{Interaction parameters for two Sodium spin mixtures. The numerical values reported here have been calculated \cite{Tiemann} in the presence of a uniform magnetic field $B=1.5$ G and have a negligible variation in the range from 0 to 2\,G.}
    \label{tab:Na-interactions}
\end{table}

\section{Static polarizability and spin-dipole oscillations}
\subsection{Static polarizability}
The great advantage of miscible mixtures with no buoyancy over buoyant or immiscible ones, relies in the fact that they allow access to the linear response of the system to small perturbations.
For example, we can directly measure the static dipole polarizability, i.e., the tendency of a two-component system to polarize as a consequence of a minimal relative displacement between the two components.  
In the clean case of equal intracomponent interactions $g_a=g_b=g$, for a vanishing displacement $x_0\rightarrow0$, the polarizability can be expressed in the linear regime using the local density approximation as
\begin{equation}
    {\mathcal{P}_0}=\frac{g+g_{ab}}{g-g_{ab}}.
\end{equation}
This expression suggests also that the static dipole polarizability is a diverging quantity, as the miscible-immiscible phase transition is approached. 

\begin{figure}[t!]
\centering
\includegraphics[width=\columnwidth]{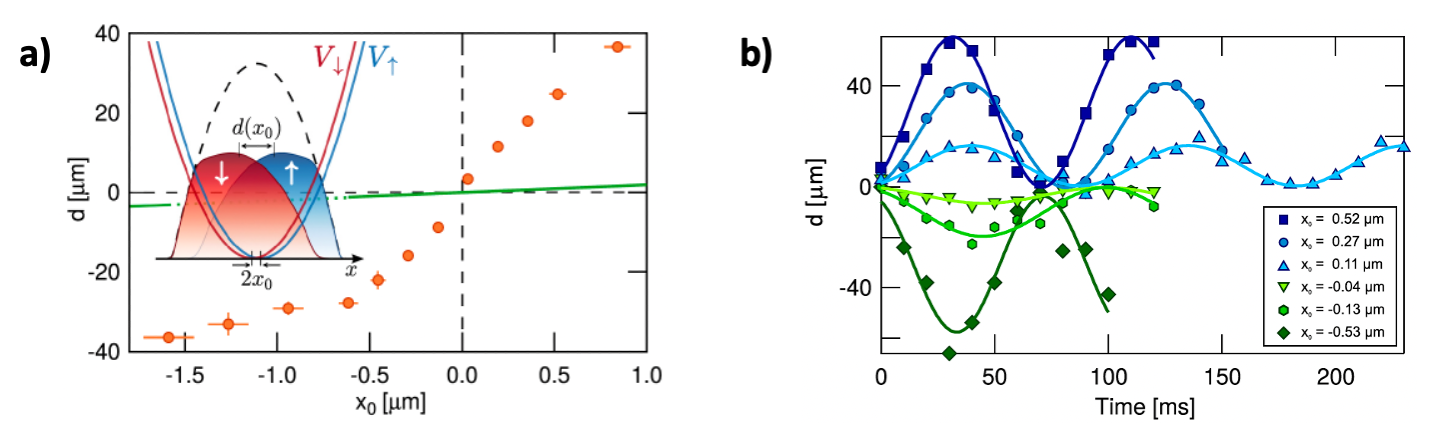}
\caption{ \textbf{a)} Center of mass displacement $d$ for different separation of the harmonic trap minima (2$x_0$) of the two components, showing the static dipole polarizability of the system $\mathcal{P}=d/2x_0$.  \textbf{b)} Spin-dipole oscillations for different amplitudes $x_0$. 
Reprinted figure with permission from T.
Bienaim\'e {\it{et al.}}, Phys. Rev. A {\textbf{94}},  063652 (2016). Copyright (2016) by the American Physical Society. }
\label{fig:fig-Polarizability}
\end{figure}

The static polarizability was directly measured in a trapped balanced Sodium mixture in $|1,\pm 1\rangle$ \cite{Bienaime2016}, by applying a small magnetic field gradient, that induces a tiny differential force on the two components of the mixture displacing their trap minima. As illustrated in the inset of Fig.~\ref{fig:fig-Polarizability}(a), for a displacement $x_0$ of each trap minimum from the equilibrium configuration, the centers of mass of the two components separate by $d$, because of the finite repulsive interaction $g_{ab}$. The polarizability of the system is experimentally evaluated as $\mathcal{P}(x_0)=d/2x_0$.  
Figure~\ref{fig:fig-Polarizability}(a) reports the experimental results from Ref.~ \cite{Bienaime2016}, showing the large response of the system for small displacements.
For comparison, the solid line corresponds to the case of no interaction between the two components ($g_{ab}=0$), for which  $\mathcal{P}=1$.

\subsection{Spin-dipole oscillation}
Another feature that this miscible mixture allows to directly observe is the spin-dipole oscillation, i.e., the oscillation of the relative displacement between the centers of mass of the two components, following an initial out-of-equilibrium configuration.  
Small amplitude spin-dipole oscillations can be excited without inducing changes in the total density.  
The relative oscillation is predicted to oscillate at the spin-dipole frequency \cite{Stringari2016}. In the limit of small displacement $\mathcal{P}_0=\mathcal{P}(0)$, it is connected to the static polarizability and to the harmonic trap frequency, through
\begin{equation}
\omega_{\rm{SD}}=\frac{\omega_x}{\sqrt{\mathcal{P}_0}}= \omega_x \sqrt{\frac{g-g_{ab}}{g+g_{ab}}}.
\label{Eq:SD}
\end{equation}

Clearly, the spin-dipole frequency becomes smaller and smaller as the miscible-immiscible transition ($g=g_{ab}$) is approached. Figure~\ref{fig:fig-Polarizability}(b) shows measurements of the spin-dipole oscillations for different initial relative velocities, taken from Ref.~\cite{Bienaime2016}. In the limit of negligible $x_0$ the prediction of Eq.~(\ref{Eq:SD}) is well confirmed by the independent measurements, connecting spin-dipole frequency and polarizability. Note that, in the case of unequal intracomponent interactions, buoyancy effects dominate and make even small amplitude oscillations become strongly anharmonic and damp out quickly, as was observed, for example, in a Rb spin mixture \cite{Hall1998} or in heteronuclear K-Rb mixtures \cite{Modugno2002,Cavicchioli2022}.

\subsection{Spin superfluidity}
\begin{figure}[t!]
\centering
\includegraphics[width=\columnwidth]{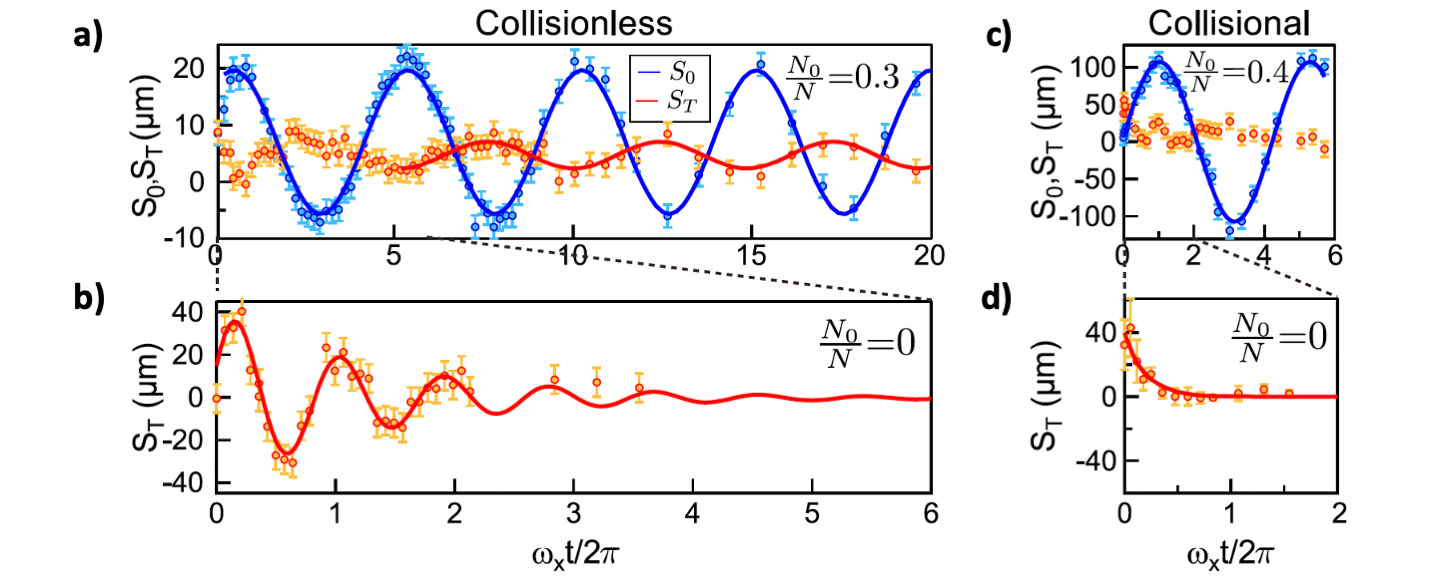}
\caption{ Relative position of the two BECs centers of mass (blue) and of the thermal components ones (red) below (\textbf{a}) and above (\textbf{b}) $T_c$, in the collisionless regime. \textbf{c-d)} Same as (\textbf{a-b}), but in the collisional regime. 
Reprinted figure with permission from 
E. Fava {\it{et al.}}, Phys. Rev. Lett. {\textbf{120}}, 170401 (2018). Copyright (2018) by the American Physical Society.}
\label{fig:fig-SpinSuperfluidity}
\end{figure}

The presence of a non-negligible thermal component partly modifies the scenario with a nontrivial dynamics involving four fluids, the BECs and the thermal components in both spin states. As reported in Ref.~\cite{Fava2018}, the two BECs can still undergo spin-dipole oscillations, with undetectable damping, while the thermal components oscillate out of phase with respect to the condensates [Fig.~\ref{fig:fig-SpinSuperfluidity}(a-b)] as a consequence of the fact that the intracomponent repulsion is stronger than the intercomponent one.
Increasing the collisional rate with a different choice of trapping parameters, the relative friction between the thermal components lead them to quickly stop their relative motion [Fig.~\ref{fig:fig-SpinSuperfluidity}(c-d)], while the BECs keep performing undamped spin-dipole oscillations, as a proof of spin superfluidity. 
In fact, if two particles with opposite spin collide, they can in general change their momentum, randomly in 3D, damping the global spin-dipole moment after many collisions. Even with elastic collisions, the only preserved quantities are the total momentum and energy. The spin flux is not preserved {\it{a priori}}. Therefore, the observation of undamped spin oscillations implies that the two BECs go through each other without friction.

Spin superfluidity was also demonstrated through the observation of a relative critical velocity, above which dissipation and magnetic vortical structures are generated \cite{Kim2017,Kim2021}.

\section{Density and spin Bogoljubov spectra}
\label{Section:Faraday}
The Bogoljubov spectrum describing the elementary excitations in a weakly interacting, single component Bose gas was measured using Bragg spectroscopy \cite{Steinhauer2022}. For a given spatial configuration of Bragg beams, the transferred energy was measured as a function of the relative frequency between the two beams allowing to find the resonant conditions and reconstruct the spectrum.
Alternatively, it is possible to measure the spectrum by modulating the trapping potential at a given frequency and observing the generation of a spatial modulation with a well definite wavelength.
The latter method was used in Ref.~\cite{Patel2020} to measure sound diffusion in a unitary Fermi gas of lithium, and recently in Ref.~\cite{Cominotti2022} to measure the excitation spectra of both density and spin channels in a balanced Sodium spin mixture, through the observation of Faraday waves.

\begin{figure}[b!]
\centering
\includegraphics[width=\columnwidth]{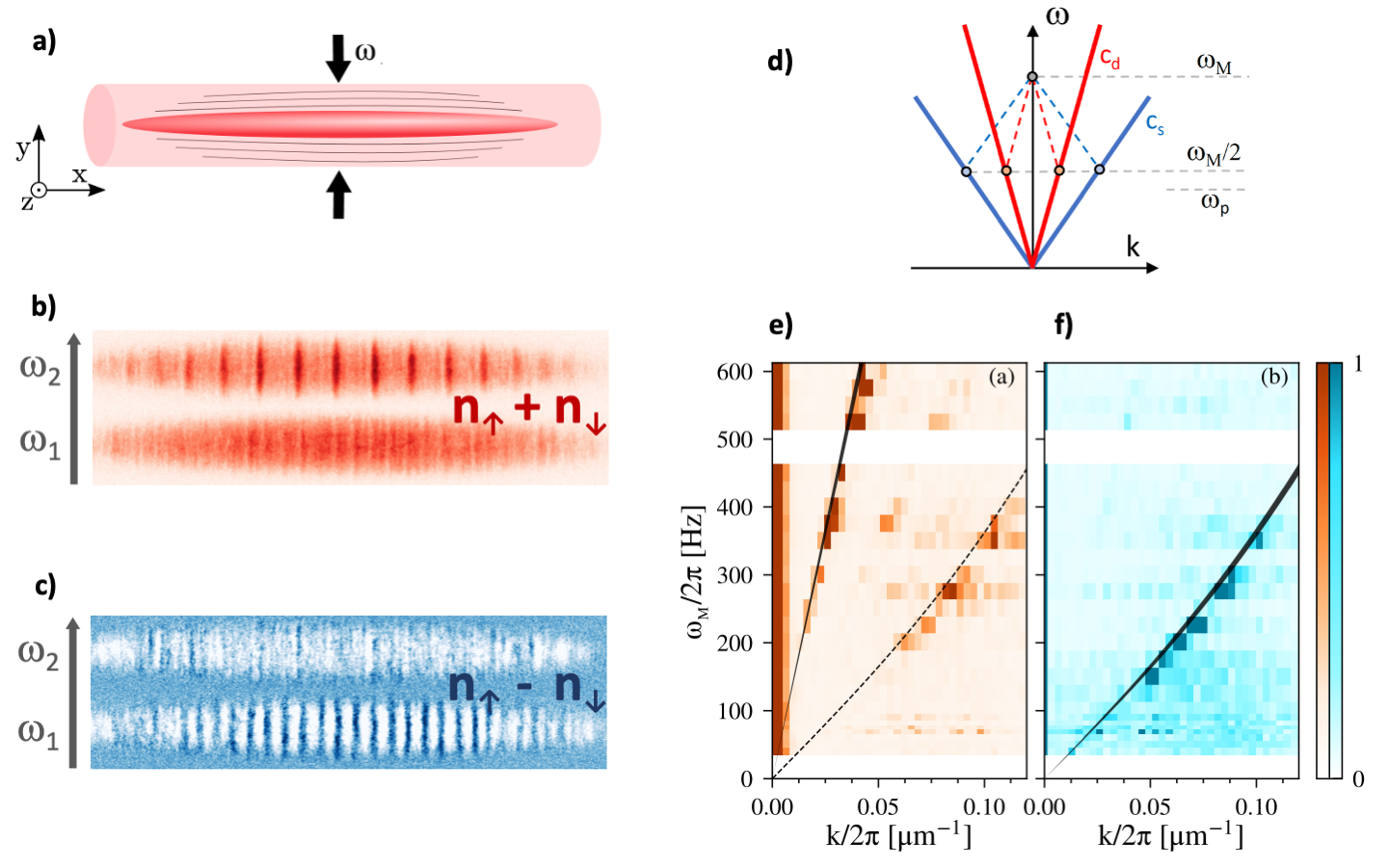}
\caption{ \textbf{a)} Sketch of the elongated mixture in the optical trap with a modulation of the radial trap frequency.  \textbf{b-c)} Total density (red) and relative density (blue) resulting by modulating at small ($\omega_1$) or higher ($\omega_2$) frequency the radial trapping confinement of the miscible Sodium spin mixture.  \textbf{d)} Decay mechanism of an excitation of energy $E=\hbar \omega_M$ in the density (red) or spin (blue) channel.  \textbf{e-f)} Comparison between predicted (lines) and measured (colorplot) Bogoljubov spectra for the density (red) and spin (blue) channels. The spin spectrum is also reported in the density one, where residual crosstalk is present. 
Reprinted figure with permission from 
R. Cominotti {\it{et al.}}, Phys. Rev. Lett. {\textbf{128}}, 210401 (2022). Copyright (2022) by the American Physical Society.}
\label{fig:fig-Faraday}
\end{figure}


As reported in Ref.~\cite{Cominotti2022}, a balanced mixture of Sodium atoms in the $|1,\pm 1\rangle$ is prepared in an elongated optical trap [see Fig.~\ref{fig:fig-Faraday}(a)]. By modulating the radial trapping frequency at $\omega_M$, the excitation with energy $E=\hbar\omega_M$ decays into pairs of longitudinal phonons with energy $\hbar\omega_M/2$ and opposite momenta. They interfere with the ground-state condensate wave function and generate a regular pattern in the mixture. Such a mechanism is resonant only with phonons in the Bogoljubov spectrum of the total density or of the spin [see Eq.~(\ref{Eq:Bogoljubov-ds})], characterized by specific wave vectors. 
Figure~\ref{fig:fig-Faraday} shows sample images of the total density (b) and of the density difference (c) for two different values of the modulation frequency, $\omega_1$ and $\omega_2$. Clear spatial patterns appear either in the total density or in the spin channel.
Figure~\ref{fig:fig-Faraday}(e-f) report the Fourier spectra of both channels for different modulation frequencies. The dark colored regions correspond to the maximum response of the system, therefore revealing the Bogoljubov excitation spectra for the two channels, independently.
The solid lines correspond to the theoretical predictions and agree very well with the experimental data.
The dashed line on the density plot of Fig.~\ref{fig:fig-Faraday}(e) corresponds to the position of the spin spectrum to explain the other peaks in the spectrum as a residual crosstalk between the two channels.

\section{Magnetic solitons}
Solitons have been realized in single component BECs \cite{Burger1999,Denschlag2000,Becker2008}, but their dynamics resulted hard to be observed on long timescales. The main reason for that relies in the typical scattering length and density values. On the one hand, a large $na$ combination is useful to ensure thermalization mechanisms within reasonable timescales, on the other hand, it makes the 
characteristic extension of the soliton ($\simeq\xi$) very small.
In most experiments, imprinted solitons have been observed decaying into other less energetic topological defects \cite{Mateo2014}, such as solitonic vortices \cite{Shomroni2009,Becker2013,Ku2014,Donadello2014,Ku2016} or vortex rings \cite{Anderson2001} .

\begin{figure}[t!]
\centering
\includegraphics[width=\columnwidth]{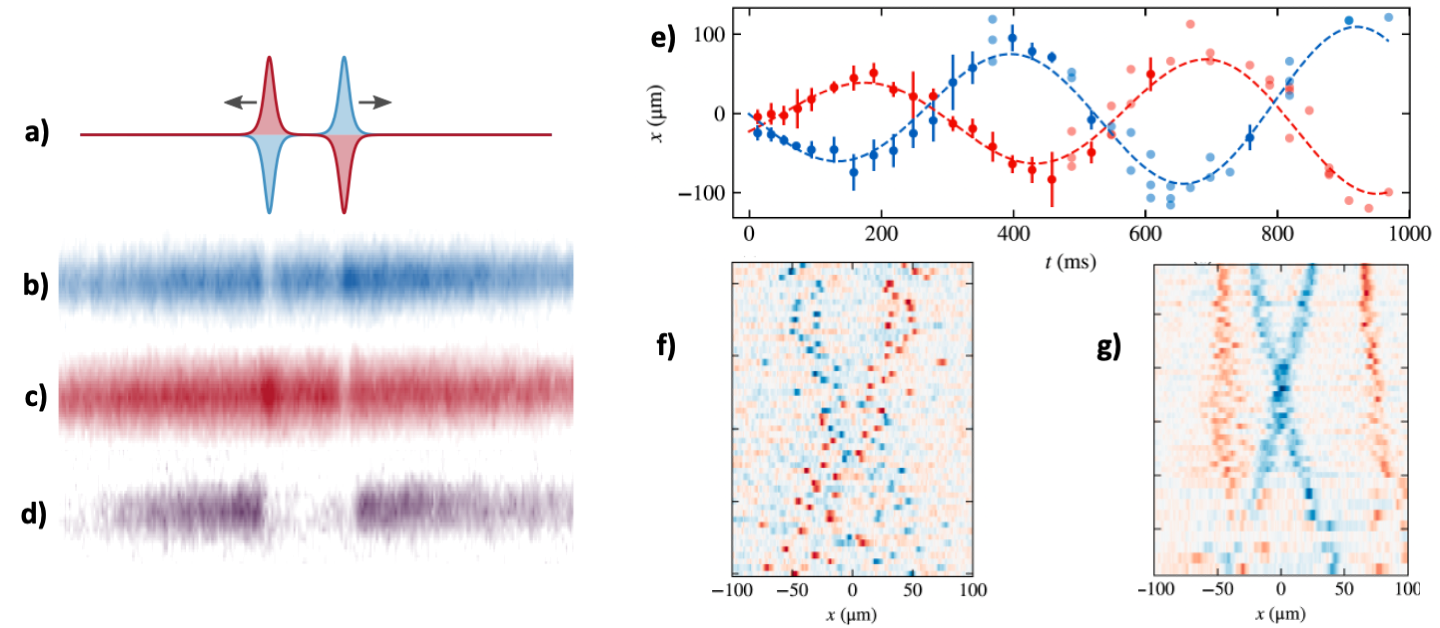}
\caption{ \textbf{a)} Sketch of two magnetic solitons moving away one from each other.  \textbf{b)} Image of component $a$ with peaks and dips at the magnetic solitons positions.  \textbf{c)} Same as b) but for component $b$, with peaks and dips inverted.  \textbf{d)} Measurement of the relative phase through interferometric techniques.  \textbf{e)} Soliton dynamics.  \textbf{f)} Collision between magnetic solitons with opposite magnetization. \textbf{g)} Same-sign soliton collision. 
Reprinted figures with permission from 
A. Farolfi {\it{et al.}}, Phys. Rev. Lett. {\textbf{125}}, 030401 (2020). Copyright (2020) by the American Physical Society.  }
\label{fig:fig-MagneticSolitons}
\end{figure}

Magnetic solitons [Fig.~\ref{fig:fig-MagneticSolitons}(a)] \cite{Qu2016,Farolfi2020,Chai2020}, instead, can be much more stable. The property of having similar interaction constants favours the condition $(g - g_{ab})\ll (g + g_{ab})$. In such a case, in fact, for a given well-measurable density of the system, the spin healing length results much larger than the corresponding density one. 

Figure~\ref{fig:fig-MagneticSolitons} reports the observation of a pair of magnetic solitons in a Sodium $|1,\pm 1\rangle$ balanced mixture \cite{Farolfi2020}. Panels b) and c) show the measured density profile of the particles in each of the two states. At the location of the magnetic soliton a density depletion in one component is accompanied by a peak in the density of the other one. Panel d) shows an interferometric measurement of the relative phase $\phi$, obtained by promoting atoms from both states to a third spin state, $|2,0\rangle$. In such a way, the full contrast observed in the outer regions is a proof of the in-phase nature of the two components, while the vanishing signal in between the two magnetic solitons is consistent with the phase jumps of $\pi$ across both magnetic solitons.

Figure~\ref{fig:fig-MagneticSolitons}(e) shows the position of the two magnetic solitons in time. They oscillate back and forth in the trap for several times as a proof of their stability. The measured oscillation period depends on the oscillation amplitude. Consistently with the prediction \cite{Qu2016}, we observe a soliton oscillation period $T=4.5\,T_x$, where $T_x$ is the period corresponding to the dipole oscillation along the axis. The oscillation amplitude slightly increases in time, as a consequence of thermal damping combined with the soliton negative effective mass.

Collisions between magnetic solitons of different and equal sign are shown in Fig.~\ref{fig:fig-MagneticSolitons}(f-g).
In the former case the two solitons can be labelled and it is clear that they pass through each other. In the latter case, they cannot be distinguished, however we can clearly see that their extension remains the same as before the collision, demonstrating their non-dispersive nature. Depending on the magnetization level, some energy might be dissipated increasing the relative momentum between the solitons after the collision, in order to avoid the excitation of the density channel, whose characteristic energy is much higher than the spin one [Eq.~(\ref{Eq:mu})].

\section{Mixture Manipulation}
\subsection{Preparation}
Balanced two-component degenerate spin mixtures can be produced either by directly cooling an unpolarized gas at a specific magnetic field that favors the desired magnetization of the final mixture \cite{JimenezGarcia2019} or by first creating a polarized condensate and then changing its internal state to form the mixture.
Following this last route, we can introduce two different protocols that produce balanced mixtures: a) $\pi$/2-pulse: switch on a high-power radiation, resonant with the energy difference between the two states, for a time corresponding to a Rabi $\frac{\pi}{2}$-pulse, b) Adiabatic Rapid Passage (ARP): adiabatically ramping the radiation detuning from large to zero, while the power is raised from zero to the final one.

The main difference between the two procedures relies in the final phase difference between the two components. 
The first method prepares  the whole system in a $\sigma_y$ state with a $\pi/2$ phase shift between the states, whereas the second method ends up in the ground state of the Hamiltonian in the presence of the coupling (along $\sigma_x$), with the two components in-phase.
However, the ARP is in general better since it is insensitive to $g_{ab}$ \cite{Farolfi2021}.

\subsection{Spin relaxation}
In a two-component mixture, three different two-body collisional processes are possible, $aa$, $bb$ and $ab$. Each of them preserves the total spin projection of the colliding atoms and, in case of elastic collisions, also the total internal energy is conserved.
However, the individual internal states before and after the collision might be different,
and the mixture undergoes spin relaxation.

If $a$ (or $b$) is in a stretched state, the spin of the system cannot relax through $aa$ (or $bb$) collisions. In the particular mixture of $|F,\pm F\rangle$ and $|F',\pm F'\rangle$, also $ab$ processes preserve the individual spins, making it a stable mixture. This is, for instance the case of Sodium atoms in $|1, -1\rangle$ and $|2, -2\rangle$, that will be studied in the following Sections.
If $a$ (or $b$) is in a non-stretched state $|F, m_F\rangle$, then an $aa$ collision can produce one atom in $|F, m_F-1\rangle$ and one in $|F, m_F+1\rangle$. In the limit of small magnetic fields, the linear Zeeman effect makes such a process preserve also the total energy. When the magnetic field is not so small, though, the second order Zeeman effect makes the total energy in the final state be different from the initial one, discerning between energetically favorable or unfavorable processes.  

If the desired two-component mixture is not naturally stable, it is still possible to inhibit spin relaxation to other (less energetic) states, by dressing them, i.e., by lifting the energy level of such states through a dedicated radiation. 
Raman or microwave radiation can be used to couple such states to other atomic states. This technique was used to study the Sodium miscible mixture $|1, \pm1\rangle$.

\subsection{Detection of density and spin}
In order to study both density and spin phenomena in ultracold mixtures it is convenient to image the two components independently and {\textit{a posteriori}} compute the sum and the difference of their optical densities at each position in space. 
When the two states of the mixture belong to the ground state manifold $F=1$, the use of an auxiliary level in $F=2$, such as $|2,0\rangle$, allows to selectively transfer atoms of either component to such a state, and consequently image those atoms illuminating them with probe light, resonant on the $F=2$ to $F'=3$ transition of the $D_2$ line. 
This choice also makes it possible to extract information on the relative phase $\phi$, by transferring both components to the auxiliary state, let them interfere and image the resulting pattern.
Microwaves with a proper polarization and frequency can be pulsed for the desired amount of time in order to transfer a controlled fraction of the gas to the auxiliary state, according to the Rabi frequency of such a transition.

Phase contrast imaging could also be implemented to have access to spatial information of both density \cite{Andrews1996} and spin \cite{Higbie2005} profiles of the mixture.

\section{Coherently-coupled mixture}
The great advantage offered by spin mixtures of a given atom, over mixtures of gases of different atomic species, consists in the possibility of state interconversion. This can be done in a controlled way using a coherent coupling radiation that connects the two internal states. The GPEs describing this coherently-coupled system are 

\begin{eqnarray}
     i \hbar \frac{\partial }{\partial t} \psi_{a} &=& \left(  - \frac{\hbar^2}{2m} \nabla ^2   +V +g_a |\psi_{a}|^2 +g_{ab} |\psi_{b}|^2 \right) \psi_{a}  -\frac{\hbar \Omega}{2}   \psi_{b}\\
      i \hbar \frac{\partial }{\partial t} \psi_{b} &=& \left(  - \frac{\hbar^2}{2m} \nabla ^2   +V - \hbar\Delta +g_b |\psi_{b}|^2 +g_{ab} |\psi_a|^2\right) \psi_b   -\frac{\hbar \Omega^*
      }{2}  \psi_{a},
\end{eqnarray}

where $\Omega$ is the strength of the coupling radiation and $\Delta$ is the frequency detuning between the driving field and the two-level system.

In this case, the stationary states are 
\begin{eqnarray}
\left(  - \frac{\hbar^2}{2m} \nabla ^2   +V +g_a |\psi_{a}|^2 +g_{ab} |\psi_{b}|^2 \right) \psi_{a}  -\frac{\hbar \Omega}{2}   \psi_{b}&=&\mu \psi_a\\
\left(  - \frac{\hbar^2}{2m} \nabla ^2   +V - \hbar\Delta +g_b |\psi_{b}|^2 +g_{ab} |\psi_a|^2\right) \psi_b   -\frac{\hbar \Omega^*
      }{2}  \psi_{a}&=&\mu \psi_b,
\end{eqnarray}

where a single, common $\mu$ appears, since only the total number of particles is a preserved quantity, while the spin is not because of the coupling radiation.

Since the two states usually belong to the same hyperfine manifold or at most to different hyperfine states of the $L=0$ level, they can be directly coupled through single photon transitions using RF or microwave radiation. Such a Rabi coupling is not associated to any measurable momentum change, i.e., the dynamics is exclusively internal. The two levels can be coupled also using two-photon transitions with the proper polarization. In this Raman coupling scheme, the net momentum transfer remains negligible when using frequencies in the RF or microwave domain, while it can be relevant if using laser radiation in the optical domain. In the latter case, the momentum transfer can range from roughly zero, for co-propagating beams, to 2$k$, for counter-propagating beams \cite{Recati2022}. Spin-orbit coupled systems will not be treated here, but pioneering experiments were reported in Refs.\cite{Lin2011}.

The internal evolution of the two-level system in the presence of the coupling radiation can be conveniently studied using the representation on the Bloch sphere. 
The state of the two-level system can be written, in general, as $|\psi\rangle = c_a |a\rangle + c_b e^{i\phi} |b \rangle$. If we consider unitary states ($|c_a|^2 + |c_b|^2=1$), then only two parameters are needed to identify the state, the population difference and the relative phase. 
By representing the population difference $Z=|c_a|^2 - |c_b|^2$ along the $z$ axis and the relative phase $\phi$ as an angle in the horizontal plane $xy$, the state is described by the unitary Bloch vector, lying on the Bloch sphere. Such a vector in the Bloch space can be considered as the spin $\bm{s}$ of the system

\begin{equation}
    \bm{s}= \left( \sqrt{1-Z^2}\cos \phi ,  \sqrt{1-Z^2}\sin \phi , Z \right).
\end{equation}

In the presence of a radiation of amplitude $\Omega$ and detuning $\Delta$ from the resonance of the two-level system, the spin  $\bm{s}$  precesses around the vector $\bm{W}=(\Omega,0,\Delta)$, at a frequency $\Omega'=\sqrt{\Omega^2+\Delta^2}$. 
If the coupling field and its detuning are kept constant, single atoms (or thermal gases with negligible interactions) change periodically their internal state in time by describing circular trajectories at constant energy around $\bm{W}$, the so-called Rabi oscillations.

\begin{figure}[t!]
\centering
\includegraphics[width=\columnwidth]{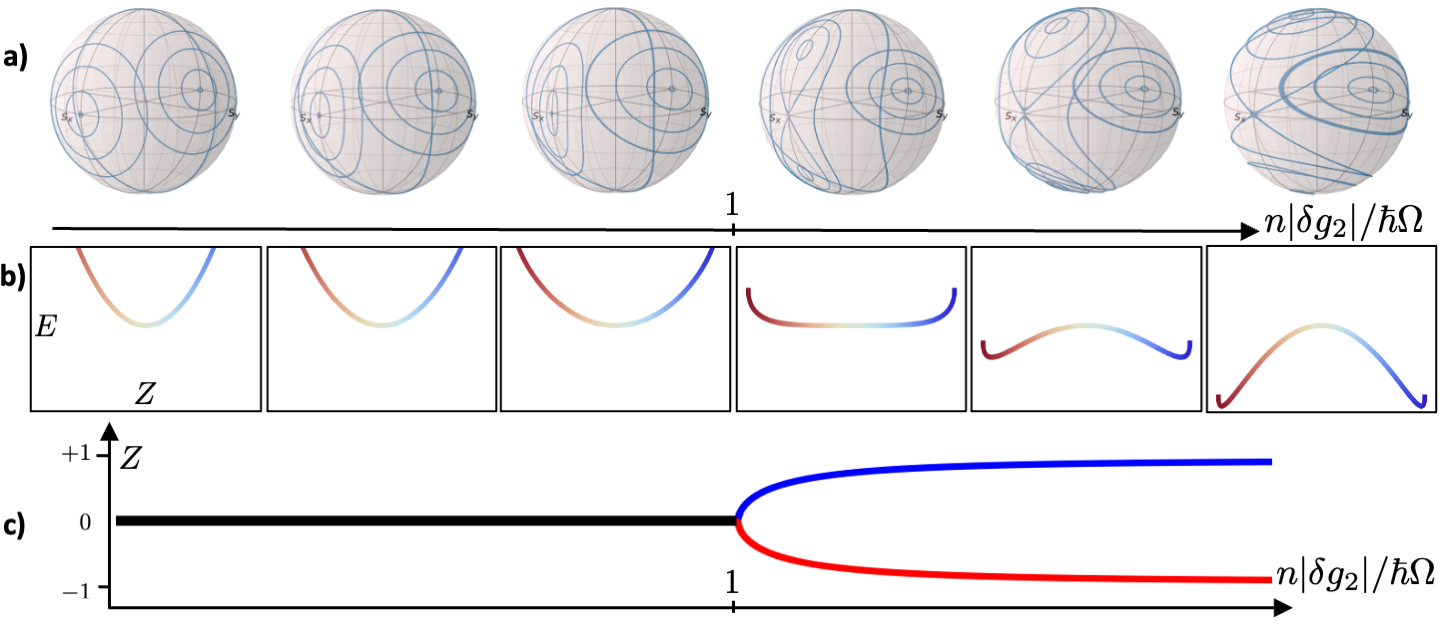}
\caption{ \textbf{a)} Bloch vector trajectories for increasing interactions in the case with $\hbar\Delta=n\delta g_1$ and $\delta g_2 <0$. \textbf{b)} Corresponding energy plot as a function of $Z$ along the arc of the Bloch sphere with $\phi=0$.  \textbf{c)} Relative magnetization $Z$ of the ground state as a function of the interactions.
The opposite case, with $\delta g_2 >0$, looks the same but after a rotation of $\pi$ around the $z$-axis and inverting the sign of the energy. The ground state, in this case, remains a single minimum even as a function of increasing interactions, but the energy maximum splits in two branches. }
\label{fig:fig-Coupling}
\end{figure}

In the case of a gas made of interacting particles, we can still use the Bloch formalism, but the modulus of the Bloch vector is no longer unitary, but rather corresponds to the density of the system, $\bm{|s|}=n$. Interactions introduce additional nonlinear terms on the $z$-component of $\bm{W}$. The modified vector is
\begin{equation}
   \bm{W}_{\mathrm{eff}}=\left( \Omega, 0, \Delta - \frac{\delta g_1\,n}{\hbar} - \frac{\delta g_2\,nZ}{\hbar} \right)
\end{equation} 
with $Z=(n_a-n_b)/(n_a+n_b)$ being the normalized magnetization and $\delta g _{1,2}=4\pi\hbar^2\delta a_{1,2}/m$ (see $\delta a_{1,2}$ in Table~\ref{tab:Na-interactions}).
The extra nonlinear terms in $\bm{W}_{\mathrm{eff}}$  make the trajectories of the Bloch vector become no longer circular, signature of a non trivial dynamics of the system [Fig.~\ref{fig:fig-Coupling}(a)]. 

If $\Delta$ is chosen to compensate for the possible asymmetry between $g_a$ and $g_b$, i.e., $\hbar\Delta=\delta g_1\,n$, then we are left with only one term along the $z$ direction, depending on the difference between the average intercomponent interaction and the intracomponent one. We can study the system as a function of the interactions $n\delta g_2 Z$.
If $\delta g_2>0$, there is always only one absolute energy minimum of the system, with a squeezing of the Bloch vector trajectories in the $z$-direction in its vicinity.
If instead $\delta g_2<0$, the mixture undergoes a $Z_2$-symmetry breaking phase transition, with a critical point determined by the coupling.  
For $n\delta g_2<\hbar\Omega$, the circular trajectories are just deformed into elliptical ones, elongated along the $z$-direction. 
At $n  \delta g_2 = \hbar\Omega$, the single minimum configuration splits into two minima for $n \delta g_2 > \hbar\Omega$. Such minima move towards the poles of the Bloch sphere, the stronger the interactions.
Figure~\ref{fig:fig-Coupling} shows how increasing interactions deform the trajectories.
In the presence of the two minima, there is one critical trajectory on the Bloch sphere (the separatrix) that separates the oscillatory dynamics around one of the minima from the outer trajectories associated to wide oscillations around both minima. For $n  \delta g_2 =2 \hbar\Omega$, the separatrix passes through the poles, therefore initially fully polarized states never invert their spin.

\subsection{Internal and external dynamics}
If the spatial extension of the system is of the order or even smaller than the spin healing length $\xi_s$, all atoms behave as a whole and the spin dynamics can be considered exclusively in the internal degrees of freedom. Such 
a dynamics was studied in detail using small Rubidium and Potassium gases and described as bosonic Josephson junctions \cite{Albiez2005,Gati2007,Zibold2010,Trenkwalder2016}. 

Instead, one can realize systems that extend for lengths that are much larger then $\xi_s$ along one or more directions. In this way, also spatial fluctuations are accessible and can be studied 
\cite{Nicklas2015,Prufer2018,Mennemann2021}.
Furthermore, extended systems can allow for a local control on the parameter $n \delta g_2 / \hbar\Omega$.
Experimentally, this can be implemented either using nonuniform confining potentials that lead to a nonuniform density $n(x)$ \cite{Nicklas2011,Nicklas2015b,Farolfi2021QT,Farolfi2021,Cominotti2022}, or  applying optical Raman coupling with nonuniform spatial profiles leading to $\Omega(x)$, or also combinations of both.

Let us focus on two different Sodium spin mixture trapped in an elongated optical potential and study the equation of state in the ground state configuration and how the system evolves, instead, when brought far from equilibrium.

The two spin mixtures under investigation (see Table~\ref{tab:Na-interactions} for their interaction properties) are: 
\begin{itemize}
    \item $|a\rangle = |2,-2\rangle$ and $|b\rangle = |1,-1\rangle$. Given its interaction properties ($\delta g_2 <0$) this mixture is an excellent candidate for the study of magnetic properties of coupled superfluid systems (see Sec.~\ref{Section:Magnetism}).
     \item $|a\rangle = |1,+1\rangle$ and $|b\rangle = |1,-1\rangle$ . The symmetry between the interactions of the two components and its miscibility make this mixture an ideal platform where to investigate elementary excitations and spin dynamics (see Sec.~\ref{Section:MassiveExcitations} and Sec.~\ref{Section:Torque}). 
\end{itemize}

\section{Coupling field stability}
It is important to note the relevance of the magnetic field stability in experiments that include the presence of a coherent coupling external radiation. In fact, any magnetic field variation happening during the experiment introduces a change in the energy difference between the two atomic states involved. The effect is equivalent to a noisy variation in time of the coupling frequency in the absence of magnetic field noise. Clearly, this would lead to quick decoherence if the amplitude of the magnetic field noise is too large. 

In order to estimate the order of magnitude of the field stability needed in these experiments with Sodium atoms, we can make considerations on the relevant energy scales. Let us start with a typical interaction energy of a standard trapped single component condensate $\mu=ng\simeq$ a few kHz. The spin interaction energy  $n\delta g$ is about two orders of magnitude smaller than $n g$, therefore around 100 Hz. 
The critical point for the magnetic transition happens on a comparable scale, when $\hbar\Omega=n\delta g_2$. If one wants to explore both sides of the transition, then, $\Omega$ should be small enough, at the level of a few tens of Hz, and correspondingly the field instability should introduce energy variations much smaller than that. In conclusion, an energy level stability of the order of a few Hz is required. This corresponds to a magnetic field stability better than a few $\mu$G.

The solution adopted by our group in Trento, to achieve such a stability, consists in surrounding the vacuum chamber with a multi-layer magnetic shield that suppresses the external field by 5 orders of magnitude \cite{Farolfi2019}. The magnetic field on the experiment is produced by a set of coils placed inside the shield, that are driven by highly stable laser drivers.

\section{Phase diagram for Magnetic systems}
\label{Section:Magnetism}
A coherently-coupled superfluid mixture has many analogies with magnetic materials, but possesses the extra features of coherence and superfluidity.
In particular, its dynamics is not affected by dissipative processes. Here, I will summarize the main analogies and the mapping between the two physical systems, and discuss the phase diagram.

The spin $\bm{s}$ of a two-level atomic gas of density $n$, dynamically evolves on the Bloch sphere in the presence of a coupling radiation with amplitude $\Omega$ and detuning $\Delta$, in a very similar way of a material with magnetic dipole  $\bm{\mu}$ and a given ferromagnetic anisotropy $J_3$, in the presence of a field $\bm{B}$. Table~\ref{tab:Magnetism} summarizes the analogies between all the terms involved.

\begin{table}[h!]
  \caption{Analogy between magnets and coherently-coupled superfluid mixtures.}
  \label{tab:Magnetism}
  \begin{tabular}{|c|c|c|c|}
    \hline
      Magnetic dipole      & Transverse field & Longitudinal field  & Ferromagnetic anisotropy\\
           $\bm{\mu}$   & $B_1$   & $B_3$ & $J_3$\\
    \hline
     spin (Bloch vector)    & Coupling   & Detuning & Interactions\\
       $\bm{s}$   & $\Omega$   & $\Delta-\delta g_1\,n/\hbar$ & $\delta g_2\,nZ/\hbar$\\
    \hline
  \end{tabular}
\end{table}

In the presence of a strong external field, the ground state of a magnet sees all dipoles aligned along the field, whereas, for a vanishing field, the ferromagnetic anisotropy dominates and the dipoles self-align along the preferential direction given by the microscopic geometry and interactions of the material.
In a very similar way, the ground state of the mixture on the Bloch sphere lies along the direction of $\bm{W}$, predominantly along $x$ in the presence of a strong external coupling field or along $\pm z$ if interactions dominate over the field.

As a magnetic dipole $\bm{\mu}$ in the presence of a magnetic field  ${\bm{B}}$ obeys the equation of motion $ \dot{\bm{\mu}}= \bm{B} \times \bm{\mu}$, the dynamics of the spin $\bm{s}$ in the Bloch space can be described through 
\begin{equation}
    \dot{\bm{s}}= \bm{W}_{\mathrm{eff}} \times \bm{s}
\end{equation}


When $\hbar\Delta=\delta g_1 n$, the mixture can undergo a para- to ferromagnetic phase transition. For 
$n \, |\delta g_2| Z < \hbar\Omega$, the system is characterized by a single energy minimum (at zero magnetization) and is paramagnetic.  For 
$n \, |\delta g_2| Z > \hbar\Omega$, instead, the system 
has two energy minima, either polarized in one state or the other, and becomes ferromagnetic.

In mixtures with $\delta g_2<0$, the quantum phase transition from para- to ferromagnet, corresponding to the bifurcation of the single minimum into a double minimum ($Z_2$ symmetry breaking), happens in the absolute ground state. Instead, the splitting happens for the energy maximum in mixtures that have $\delta g_2>0$.

\begin{figure}[t!]
\centering
\includegraphics[width=\columnwidth]{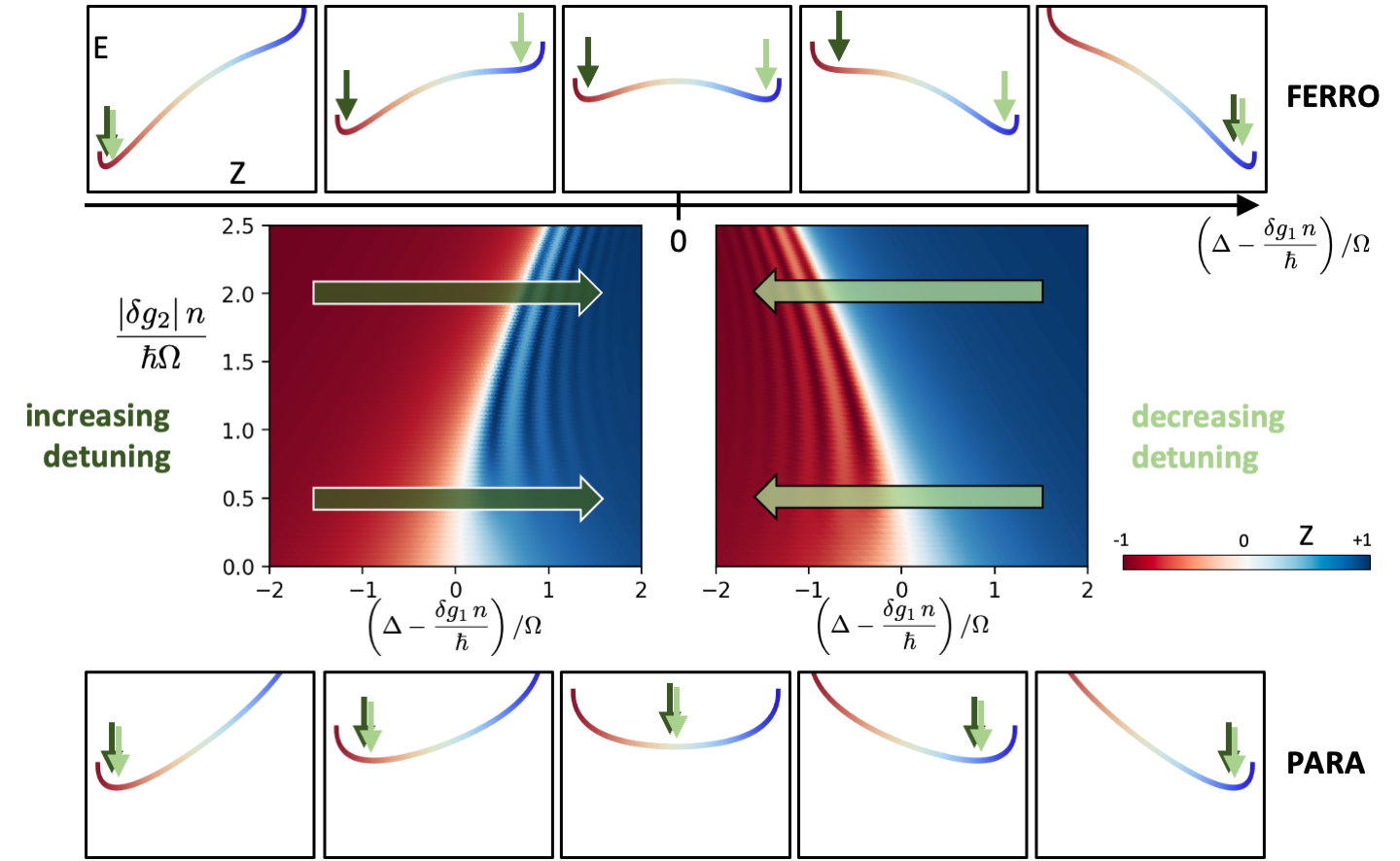}
\caption{The small panels, above and below, show the energy diagrams for systems in the ferromagnetic regime
($|\delta g_2|\,n/\hbar\Omega=2$) and in the paramagnetic one ($|\delta g_2|\,n/\hbar\Omega=0.5$), for different values of $(\Delta -\delta g_1\,n/\hbar)/\Omega$. 
In the paramagnetic case, there is always a single minimum and it follows slowly the change of the detuning.
In the ferromagnetic case, two energy minima are present around the resonant condition.  
Starting with a large negative detuning in the red state and increasing the effective detuning (see the dark green arrows), the system stays in the local red minimum until such red minimum disappears and the system jumps to the absolute blue minimum. The small vertical arrows indicate the position of the local minimum where the system is, following the ramp with increasing detuning.
Starting with a large positive detuning and reversing the direction of the ramp (light green arrow) shows a different behavior in the ferromagnetic region. 
The central panels show the magnetization of the system, after a ramp on the effective detuning. }
\label{fig:fig-Hysteresis}
\end{figure}

We can investigate the para-to-ferromagnetic phase transition using the immiscible mixture. 
In order to avoid confusion between the coordinates in real space and in the spin space, we can label them as 123 and $xyz$, respectively. Let us, then, consider an atomic system, extended along direction 1. 
In the experiment, we can start with a fully polarized gas in $|1,-1\rangle$. By switching on an initially far-detuned coupling and then adiabatically reducing it, realizing the so-called adiabatic rapid passage (ARP). The Bloch vector, in that case, follows the variation of $\bm{W}_{\mathrm{eff}}$ and by stopping the ramp in $\Delta$, the system remains in its state in the presence of the coupling $\Omega$.

Given the interaction parameters $\delta g_{1,2}$, the coupling $\Omega$ the detuning $\Delta$, the energy of the system can be written in terms of the relative magnetization $Z$ and the azimuthal angle on the Bloch sphere $\phi$ as
\begin{equation}
      E(Z,\phi) \propto \hbar \Delta - \delta g_1 n Z - \frac {\delta g_2 n}{2} Z^2 - \hbar \Omega \sqrt{1-Z^2}\cos \phi.
      \label{eq:energy1}
 \end{equation}
Figure~\ref{fig:fig-Hysteresis} shows the energy profiles for $\phi=0$ as a function of $Z$ for different values of $(\Delta-\delta g_1 n /\hbar)$.

Ramping the effective detuning across the critical region in opposite ways provides different results, as shown in the central plots in Fig.~\ref{fig:fig-Hysteresis}. Such a difference is a direct evidence of the hysteresis characterizing the para- to ferromagnetic transition. In fact, the presence of the double minimum in the ferromagnetic region gives the system the possibility to remain in the local minimum, even if it is no longer the absolute one for some parameter range.
A recent experiment in Trento \cite{Cominotti2023} used the Sodium immiscible mixture made of $|2,-2\rangle$ and  $|1,-1\rangle$ trapped in an elongated harmonic trap to reconstruct the full phase diagram across the para- to ferromagnetic phase transition, through the direct observation of an interaction-dependent  hysteresis. There, the inhomogeneous density distribution was used to have at once both paramagnetic and ferromagnetic regions in a given sample.

\section{Massive many-body elementary excitations}
\label{Section:MassiveExcitations}

In the presence of coherent coupling the excitation spectra [Eq.~(\ref{Eq:Bogoljubov-ds})] change, acquiring a gap and turning from linear to quadratic for small $k$. 
In particular, the Bogoljubov spectrum for the density channel remains unaltered, while the spin one becomes
\begin{equation}
    E_{s}(k,\Omega) = \hbar \omega_{s}(k,\Omega) = \sqrt{\left(\frac{\hbar^2 k^2}{2m}+\hbar\Omega\right)\left( \frac{\hbar^2 k^2}{2m}+2mc_{s}^2 +\hbar\Omega\right)}.
    \label{Eq:Bogoljubov-Omega}
\end{equation}

The energy gap $E_0$ depends on $\Omega$ and can be easily obtained from Eq.~(\ref{Eq:Bogoljubov-Omega}) for $k=0$, $E_0(\Omega)=\sqrt{\hbar\Omega(\hbar\Omega+2mc_s)}$. Such a gapped spectrum can be interpreted as a massive excitation spectrum $E_{s}(k,\Omega)=E_0(\Omega)+ \frac{\hbar^2k^2}{2M(\Omega)}$, with an effective mass of the excitation $M(\Omega)=2mE_0\hbar\Omega/(E_0^2+\hbar^2\Omega^2)$.

Using the same method described in Sec.~\ref{Section:Faraday}, the Bogoljubov spectrum for the coherently coupled case was measured through the excitation of Faraday spin waves \cite{Cominotti2022}.

\begin{figure}[b!]
\centering
\includegraphics[width=\columnwidth]{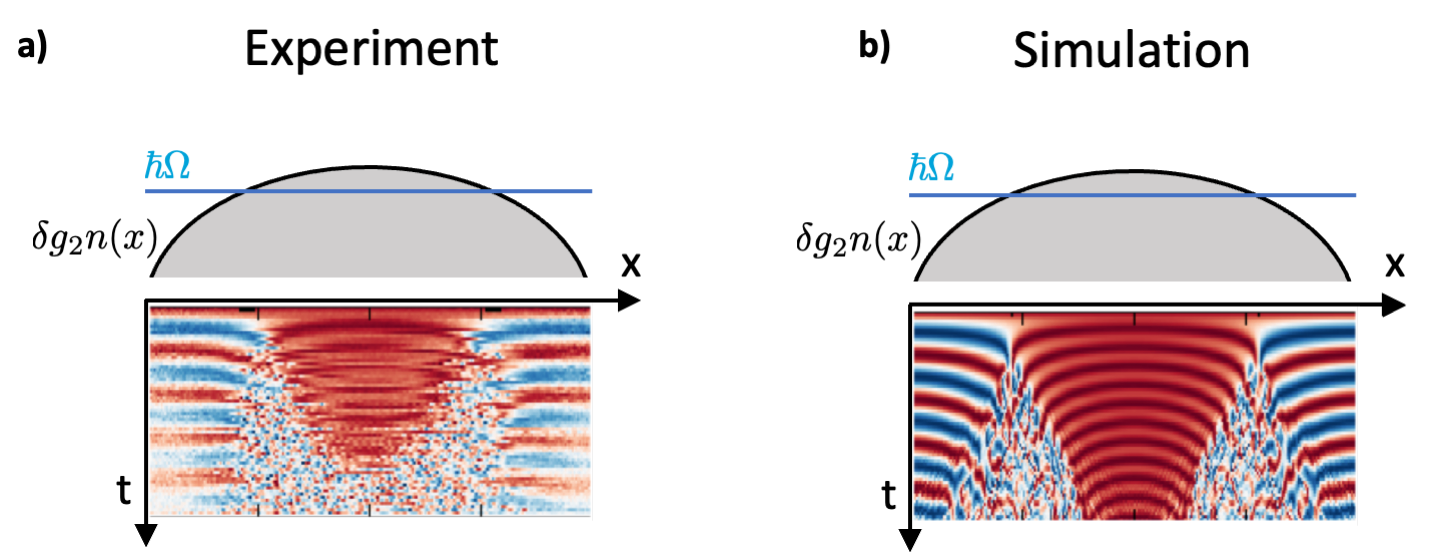}
\caption{ \textbf{a)} Experimental observation of the spin dynamics for an initially fully polarized extended system (red) in the presence of coherent coupling. First inner and outer regions show a different evolution, then highly magnetized spin waves are generated at the interfaces. \textbf{b)} GPE simulation of the same system. Figure adapted from A. Farolfi {\textit{et al.}}, Nature Physics {\textbf{17}}, 1359 (2021). Copyright © 2021, Springer Nature. }
\label{fig:fig-QuantumTorque}
\end{figure}


\section{Far-from-equilibrium spin dynamics}
\label{Section:Torque}

We have seen that extended systems allow to have different regions with different magnetic properties, depending on the local balance between interactions and coupling field.
Let us now investigate the nontrivial dynamics of extended inhomogeneous systems. In particular let us focus on the case of a miscible Sodium mixture. As mentioned above, the energy profiles are the same as in the immiscible case, but with an inversion between the ground and excited state. So, the ground state has now always a single minimum.
If we prepare the system in a state near such energy minimum, the corresponding Bloch vector precesses around the minimum with a frequency $\Omega'\simeq\sqrt{\Omega^2+\Delta^2}$ \cite{Farolfi2021}. 
If we prepare the system in a far-from-equilibrium initial condition, it then evolves in a non trivial way. The extended nature of the sample adds on top of the internal dynamics, also contributions related to the spin current.
On short timescales it follows the local conditions, but soon the spin current will start playing a non-negligible role.

Figure~\ref{fig:fig-QuantumTorque}(a) shows the dynamics of a fully polarized elongated system in $|1,-1\rangle$ after switching on a (two-photon, microwave) coherent coupling with the state $|1,+1\rangle$. In this case $\Delta=0$ and $\Omega$ is smaller than $n\delta g_2/\hbar$ in the central region and larger on the outer ones, because of the TF density profile. Clearly, in the central part, where interactions dominate, the system remains polarized, whereas in the outer regions, where the external field dominates, the system undergoes full Rabi oscillations between the two states \cite{Farolfi2021QT}. 
Globally, such an extended superfluid system can be considered as a dissipationless 1D magnetic heterostructure with alternating para-ferro-para magnetic-like materials.

At the interfaces between the two distinct magnetic materials
the magnetization accumulates fast within a small length and the system cannot support such a strong magnetization gradient causing the breaking of the interfaces and the generation of spin waves that propagate in the system. The large difference between the density and spin energy scales, allows to explore the spatial dynamics purely given by the spin current, with a negligible contribution of the density one. In this case the coupled superfluid atomic system can be described using the dissipationless Landau-Lifshitz equations \cite{Farolfi2021QT}.

\section{Conclusions}
In this Lecture, we have revisited the basic properties of superfluid mixtures, from the ground state configuration to the elementary excitations, from collective modes to topological excitations. A special focus was dedicated to two specific two-component spin mixtures of Sodium atoms, a perfectly symmetric and miscible one and an immiscible one, with and without coherent coupling. They were used as ideal platforms for the investigation of elementary excitations \cite{Bienaime2016,Fava2018,Cominotti2022,Farolfi2020} and for the study of dissipationless para- to ferromagnetic phase transition \cite{Farolfi2021,Farolfi2021QT,Cominotti2023}. 
Such spin mixtures can be excellent candidates for spintronic applications, but they can also
employed to investigate phenomena that have strong analogies to those in high-energy physics of cosmology. For example, confinement of half quantum vortices \cite{Seo2015} in a coherently-coupled mixture is expected to happen, and show a strong connection to quark confinement \cite{Son2002,Eto2018}. Furthermore, corotating spin mixtures could be employed to explore the region near the vortex core, were spin fluctuations are expected to be observable in the ergoregion (distances smaller than $\xi_s$ and larger than $\xi_d$) \cite{Berti2023}.

\acknowledgments

The author thanks Gabriele Ferrari, Alessio Recati and Alessandro Zenesini, for their critical reading of the manuscript, and all the members of the Pitaevskii BEC Center in Trento for the stimulating discussions and fruitful collaboration. Finantial support is acknowledged from Provincia Autonoma di Trento and from the European Union Horizon 2020 research and innovation Programme through the STAQS project of QuantERA II (Grant Agreement No. 101017733).

\bibliographystyle{varenna}
\bibliography{bibliography_Varenna}

\end{document}